\documentclass{jfm}

\pdfoutput=1

\input{jfmdefs}

\usepackage{amsmath,dsfont} 
\usepackage{bm}
\usepackage{subfigure}
\usepackage{graphicx}
\usepackage{natbib}
\newcommand{\url}[1]{\texttt{#1}}
\graphicspath{{figs/}}

\newcommand{\mathnotation}[2]{\newcommand{#1}{\ensuremath{#2}}}

\renewcommand{\l}{\left}
\renewcommand{\r}{\right}
\renewcommand{\time}{t}
\mathnotation{\ldef}{\mathrel{\raisebox{.069ex}{:}\!\!=}}
\mathnotation{\rdef}{\mathrel{=\!\!\raisebox{.069ex}{:}}}
\mathnotation{\ee}{\mathrm{e}}              

\mathnotation{\dint}{\,{\mathrm{d}}}        
\mathnotation{\xv}{\bm{x}}                  
\mathnotation{\urh}{u_\rho}                 
\mathnotation{\uz}{u_z}                     
\mathnotation{\ruv}{\hat{\bm{r}}}           
\mathnotation{\ptl}{\partial}               
\mathnotation{\Uc}{U}                       
\mathnotation{\pal}{\lambda}                
\mathnotation{\Dist}{L}                     
\mathnotation{\lsc}{\ell}                   
\mathnotation{\nd}{n}                       
\mathnotation{\ai}{a}                       
\mathnotation{\bi}{b}                       
\mathnotation{\dadb}{\dint\ai\dint\bi}      
\mathnotation{\da}{\dint\ai}                
\mathnotation{\db}{\dint\bi}                
\mathnotation{\Vol}{V}                      
\mathnotation{\Nenc}{M}                     
\mathnotation{\wake}{\text{wake}}           
\mathnotation{\Deltarvar}{\delta}           
\mathnotation{\C}{C}                        
\mathnotation{\Rdisk}{R}                    

\title[Stirring by squirmers]{Stirring by squirmers}

\author[Z. Lin, J.-L. Thiffeault, and S. Childress]%
{
  Z\ls H\ls I\ns
  L\ls I\ls N$^1$,\ns
  J\ls E\ls A\ls N\ls -- L\ls U\ls C\ns
  T\ls H\ls I\ls F\ls F\ls E\ls A\ls U\ls L\ls T$^{1,2}$, \\
\and
  S\ls T\ls E\ls P\ls H\ls E\ls N\ns
  C\ls H\ls I\ls L\ls D\ls R\ls E\ls S\ls S$^3$
}

\affiliation{
  $^1$ Institute for Mathematics and Applications, University of
  Minnesota -- Twin Cities, 207 Church Street S.E., Minneapolis, MN
  55455, USA\\
  $^2$ Department of Mathematics, University of Wisconsin -- Madison,
  480 Lincoln Drive, Madison, WI 53706, USA\\
  $^3$ Courant Institute of Mathematical Sciences, \hbox{New York
  University}, 251 Mercer Street, \hbox{New York}, NY 10012, USA
}

\begin{document}

\maketitle

\begin{abstract}
  We analyse a simple `Stokesian squirmer' model for the enhanced
  mixing due to swimming micro-organisms.  The model is based on a
  calculation of Thiffeault \& Childress [\emph{Physics Letters A}
  \textbf{374}, 3487 (2010)], where fluid particle displacements due
  to inviscid swimmers are added to produce an effective diffusivity.
  Here we show that for the viscous case the swimmers cannot be
  assumed to swim an infinite distance, even though their total mass
  displacement is finite.  Instead, the largest contributions to
  particle displacement, and hence to mixing, arise from random
  changes of direction of swimming and are dominated by the far-field
  stresslet term in our simple model.  We validate the results by
  numerical simulation.  We also calculate nonzero Reynolds number
  corrections to the effective diffusivity.  Finally, we show that
  displacements due to randomly-swimming squirmers exhibit PDFs with
  exponential tails and a short-time superdiffusive regime, as found
  previously by several authors.  In our case, the exponential tails
  are due to `sticking' near the stagnation points on the squirmer's
  surface.
\end{abstract}

\section{Introduction}

Swimming creatures affect their environment in many ways, and one
which has received attention recently is how they mix the
surrounding fluid.  This phenomenon is called biogenic mixing or biomixing.
The most striking and controversial setting is the ocean:
\citet{Dewar2006} suggested that marine life might have an impact on
vertical mixing in the ocean.

\citet{Katija2009} proposed that the dominant effect involved in
biomixing is the mass displacement due to a swimming body.  This
phenomenon is called Darwinian drift, after~\citet{Darwin1953}, though
the displacement due to a moving cylinder was obtained
by~\citet{Maxwell1869}.  \citet{Thiffeault2010b} derived the effective
diffusivity of an `ideal gas' of randomly-distributed non-interacting
swimmers, and showed that it depends on the induced
squared-displacement of fluid particles by the swimmers, as opposed to
the net mass displaced.

In the present paper we apply the techniques of \citet{Thiffeault2010b} to
swimmers in the low-Reynolds number regime.  Since the experiments of
\citet{Wu2000}, there has been considerable work in that direction.  Two
general types of swimmers arise: pushers or pullers, depending on whether the
propulsion takes place ahead or behind the swimmer's centre of mass.  Several
authors~\citep{Dombrowski2004,
  HernandezOrtiz2006,Saintillian2007,Underhill2008} have pointed out that at
high volume fractions the pushers align spontaneously with their neighbours,
producing large-scale motions.  At low volume fractions, pushers and pullers
behave similarly, moving in a relatively uncorrelated manner.  It is this
limit that we treat in this paper.  We use \citet{Lighthill1952} and
\citet{Blake1971}'s `squirmer' model, where a swimmer is modelled as a sphere
in Stokes flow with a prescribed tangential velocity distribution.  These have
recently been studied by~\citet{Ishikawa2007b} and~\citet{Drescher2009} as
models of the green alga~\emph{Volvox}.  Squirmers could also be appropriate
for \emph{Chlamydomonas}~\citep{Leptos2009, Guasto2010, Drescher2010}, as the
observed velocity fields resemble those of the models described here.  We find
a formula for the effective diffusion coefficient due to the squirmers, and
show that it depends predominantly on the far-field displacements imposed by
the swimmer.  Moreover, the dominant contributions arise from the turning
motion of the swimmers, in sharp contrast to the potential flow case
\citep{Thiffeault2010b} where the path length of straight swimming could be
considered infinite for all practical purposes.  We also show that the
probability distribution function of displacements due to squirmers has
exponential tails, as in \citet{Leptos2009}, and that it exhibits a short-time
superdiffusive regime as in \cite{Wu2000}.

The formula we derive for the effective diffusivity has a wide applicability
beyond squirmers, for instance to swimmers at larger Reynolds numbers with
wakes.  In that spirit, we derive finite (but small) Reynolds number
corrections to the effective diffusivity using the Oseen form for the far
field, and find that the effective diffusivity falls off as $\Rey^{-0.61}$
from its Stokes (zero $\Rey$) value.

\section{Computing the effective diffusivity}

\begin{figure}
  \centering
  \includegraphics[width=.55\columnwidth]{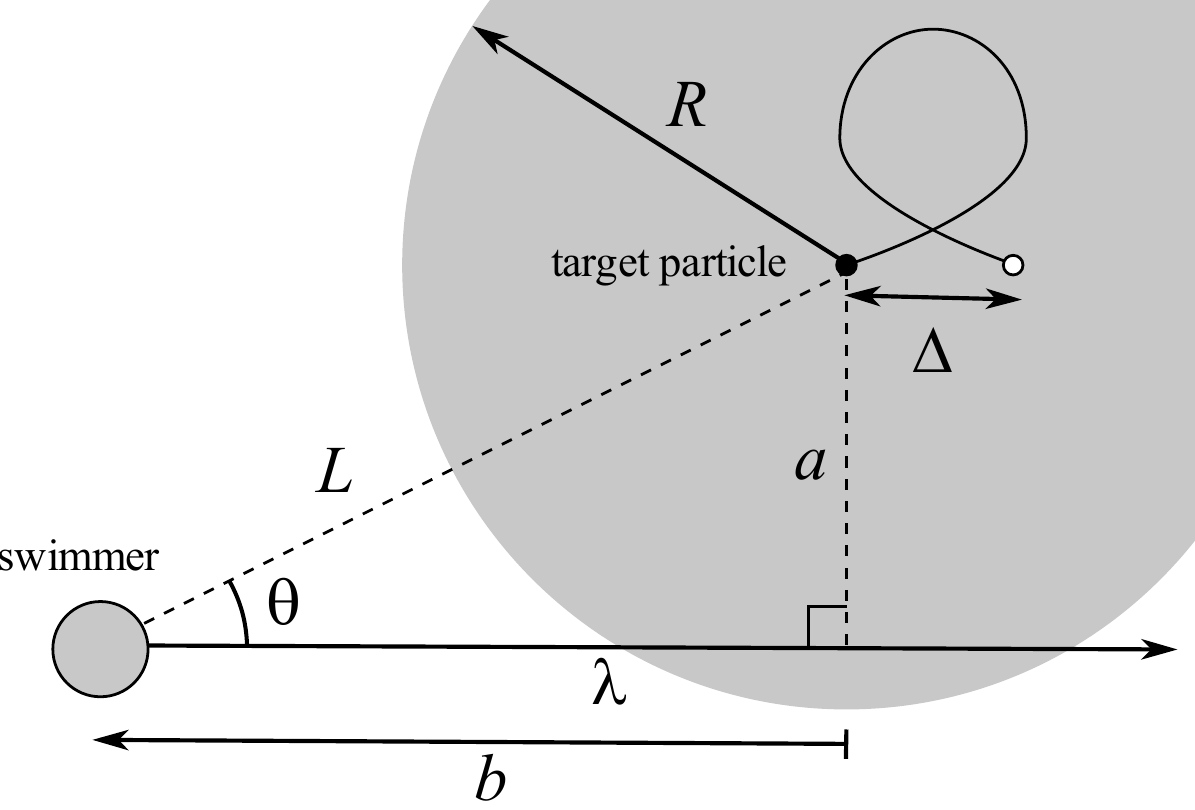}
  \caption{Definition of impact parameters~$\ai$ and~$\bi$,
    displacement~$\Delta=\Delta_\pal(\ai,\bi)$, and swimming path
    length~$\pal$.  In this picture the parameter~$\bi$ is positive;
    negative~$\bi$ corresponds to the swimmer starting its trajectory
    past the point of initial closest approach.  The filled
    dot is the initial position of the target particle and the hollow
    dot is its final position after the swimmer has moved by a
    distance~$\pal$.  The `interaction disk' of radius~$\Rdisk$ is
    also shown.}
  \label{fig:abdiagram}
\end{figure}

We first consider a single swimmer in two or three dimensions that moves in a
straight line at constant speed~$\Uc$ for a distance~$\pal$.  For simplicity,
the swimmer is assumed axially symmetric with characteristic size~$\lsc$, and
moves along its axis of symmetry.  We are concerned with the effect of the
swimmer on a target Lagrangian particle.  There are two `impact parameters'
necessary to describe such a situation: the initial perpendicular distance to
the particle, $\ai \ge 0$, and the relative distance~$\bi$ of the start of the
trajectory to the point of initial closest approach (see
figure~\ref{fig:abdiagram}).  Note that $\bi$ may be positive or negative, and
is positive if closest approach would occur at a positive time.
After the swimmer traverses its full path length~$\pal$, the target particle
is displaced by a distance~$\Delta_\pal(\ai,\bi)$.

Inspired by Einstein's theory of Brownian motion~\citep{Einstein1905}
and following~\cite{Thiffeault2010b}, we now model the
displacement of the target particle after~$\Nenc$ encounters with
swimmers as a series of `kicks', each of which moves the particle
by~$\Delta_\pal(\ai_k,\bi_k)$ in some random direction~$\ruv_k$.  The
position~$\xv_\Nenc$ of the particle is
\begin{equation}
  \xv_\Nenc = \xv_0 + \sum_{k=1}^\Nenc \Delta_\pal(\ai_k,\bi_k) \, \ruv_k
\end{equation}
and we will assume~$\xv_0=0$.  The swimmers are assumed identical and
non-interacting, and the encounters are taken to be statistically
independent.  Their directions of swimming are assumed isotropically
distributed.  In this case, the mean square displacement after $\Nenc$
such encounters is given by
\begin{equation}
  \<\lvert\xv_\Nenc\rvert^2\>
  = \sum_{k=1}^\Nenc \<\Delta_\pal^2(\ai_k,\bi_k)\>
  = \Nenc \<\Delta_\pal^2(\ai,\bi)\>,
  \label{eq:meansq}
\end{equation}
because the~$\ai_k$ and~$\bi_k$ are identically distributed for each
encounter, with the brackets denoting an average over~$\ai$ and~$\bi$.  (We
could also average over distributions of~$\pal$ and~$\Uc$, but we assume
constant values for simplicity.)  Our theory will apply in the dilute limit,
which we will define more carefully below.

The displacement $\Delta_\pal$ is assumed small compared to $\pal$, so
that we can regard the target particle as approximately fixed during
an encounter.  We assume that each swimmer advances by a fixed
length~$\pal$ and then changes direction randomly, so after a time
$\time$, $\Uc\time/\pal$ such direction changes have occurred.  For
the 2D case, the number of swimmers in a ring of radius~$\Dist$ and
thickness~$\dint\Dist$ around the target particle
is~$2\pi\nd\Dist\dint\Dist$, where~$\nd$ is the number density of
swimmers.  To define an encounter, we introduce an `interaction disk'
(sphere in 3D) of radius~$\Rdisk$ around the target particle (see
figure~\ref{fig:abdiagram}), and consider an encounter as occurring
when a swimmer enters that disk.  If a swimmer does not enter the
interaction disk, then the motion of the target particle is
negligible.  The interaction disk allows us to be
more precise about what we mean by a dilute suspension of swimmers: we
require $\nd\Rdisk^2\ll 1$ ($\nd\Rdisk^3\ll 1$ in 3D) so that
encounters are isolated from each other.

The expected number~$\Nenc$ of such
encounters after a time~$\time$ is best written in terms of integrals
over~$\ai=\lvert\Dist\sin\theta\rvert$ and~$\bi=\Dist\cos\theta$:
\begin{equation}
  \Nenc = \frac{2\nd\Uc\time}{\pal}\,
  \int_0^\Rdisk
  \int_{-\sqrt{\Rdisk^2-\ai^2}}^{\pal+\sqrt{\Rdisk^2-\ai^2}}\db\da
  = 2\nd \Rdisk\Uc\time + \pi\nd \Uc\time\,\frac{\Rdisk^2}{\pal}\,.
\end{equation}
For large path length~$\pal\gg \Rdisk$ the final term can be
neglected, and we obtain the same expression as
in~\cite{Thiffeault2010b}, where an infinite path length was assumed:
for large path length, the frequency of encounters is independent
of~$\pal$.

Next we compute the average squared displacement~\eqref{eq:meansq} by
integrating~$\Delta_\pal^2$:
\begin{equation}
  \Nenc \<\Delta_\pal^2\> = \frac{2\nd\Uc\time}{\pal}\,
  \int_0^\Rdisk
  \int_{-\sqrt{\Rdisk^2-\ai^2}}^{\pal+\sqrt{\Rdisk^2-\ai^2}}
  \Delta_\pal^2(\ai,\bi) \db\da\,.
\end{equation}
We assume that~$\Delta_\pal(\ai,\bi)$ decays rapidly for
large~$\ai\gg\lsc$, where~$\lsc$ is the characteristic size of a
swimmer.  In that case we can replace~$\Rdisk\gg\lsc$ by infinity;
after also replacing~$\xv_\Nenc$ by~$\xv(\time)$ we obtain
\begin{equation}
  \<\lvert\xv(\time)\rvert^2\> =
  \frac{2\nd\Uc\time}{\pal}\int_0^\infty \int_{-\infty}^{\infty}
  \Delta_\pal^2(\ai,\bi)\db\da
  \rdef 4\kappa \time
  \quad ({\rm 2D}),
  \label{eq:main2d}
\end{equation}
where we introduced the effective diffusivity~$\kappa$.  In 3D the
analogous expression can easily shown to be
\begin{equation}
  \<\lvert\xv(\time)\rvert^2\> =
  \frac{2\pi\nd\Uc\time}{\pal}\int_0^\infty \int_{-\infty}^{\infty}
  \ai\,\Delta_\pal^2(\ai,\bi)\db\da
  \rdef 6\kappa \time
  \quad ({\rm 3D}).
  \label{eq:main3d}
\end{equation}
Note that the `interaction disk' of radius~$\Rdisk$ has disappeared
from~\eqref{eq:main2d} and~\eqref{eq:main3d}: because of the rapid
decay of $\Delta_\pal(\ai,\bi)$ with~$\ai$, we are free to overcount
encounters when computing~$\<\lvert\xv(\time)\rvert^2\>$ because
faraway encounters hardly displace the target particle.

\cite{Thiffeault2010b} discussed models based upon potential flow past
a cylinder or a sphere and made use of an approximate form
of~\eqref{eq:main2d}--\eqref{eq:main3d}.  The approximate form
effectively assumes that the particle doesn't move unless the swimmer
actually crosses the point of initial closest approach (a good
approximation for potential flow, see
figure~\ref{fig:sphere_dispint}).  Thus $\Delta_\pal(\ai,\bi)$ takes
the `top-hat' form
\begin{equation}
  \Delta_\pal(\ai,\bi) =
  \begin{cases}
    \Delta(\ai),&\text{if $0 < \bi < \pal$,}\\
    0,&\text{otherwise.}
  \end{cases}
  \label{eq:Deltahat}
\end{equation}
This approximation works very well for potential flow.  With the
approximate form~\eqref{eq:Deltahat}, the effective diffusivities
defined by~\eqref{eq:main2d}--\eqref{eq:main3d} become
\begin{equation}
  \kappa =
  \begin{cases}
    \frac{1}{2} \Uc \nd
    \int_0^\infty \Delta^2(\ai)\da,&\text{(2D)},\\
    \frac{\pi}{3}\Uc \nd
    \int_0^\infty \ai\Delta^2(\ai)\da,&\text{(3D)},
\end{cases}
\label{eq:tophat}
\end{equation}
as in \citet{Thiffeault2010b}.  Because the potential flow
approximation is inappropriate for micro-organisms, we will not
use~\eqref{eq:tophat} but rather take the full
form~\eqref{eq:main2d}--\eqref{eq:main3d}.

Note that the number density scaling
of~\eqref{eq:main2d}--\eqref{eq:main3d} was derived
by~\cite{Underhill2008} using the Green--Kubo formula, but here we get
explicit values for the prefactor.  As we will see below, this allows
us to determine where the dominant contribution to~$\kappa$ arises
from.  In addition, the dimensionless prefactor can vary significantly
depending on the swimmer under consideration.

\section{Stirring by squirmers}

\emph{Squirmers} are a simple model of swimming micro-organisms in the
Stokesian regime.  \cite{Lighthill1952} first introduced the model and
\cite{Blake1971} later extended the analysis.  Recently, the squirmer
model has been applied to describe the behaviour of slow-moving
ciliates \cite[see for example][]{Ishikawa2006, Ishikawa2007b,
  Drescher2009}, whose motion is propelled by arrays of pulsating
cilia on their surface.  A squirmer is generally a three-dimensional
sphere moving in a Stokesian fluid with velocity field prescribed on
its boundary.  The velocity field generated by such a squirmer is
derived by an expansion of solutions obtained by separation of
variables.  The steady swimming imposes a force-free condition between
the fluid and the body which in turn translates to algebraic equations
for the expansion coefficients of the solution.

\begin{figure}
  \centering
  \includegraphics[width=.75\columnwidth]{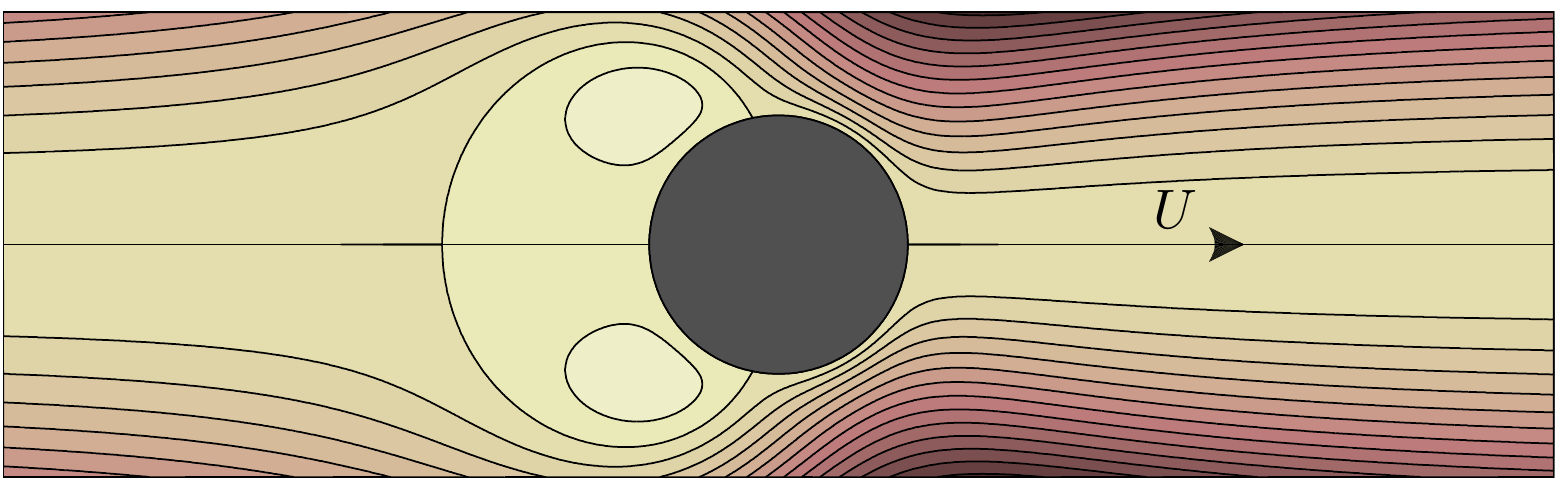}
  \caption{Streamfunction contours for a steady squirmer
    with~$\beta=5$, in a comoving reference frame.  Note the wake or
    closed `bubble' behind the squirmer.}
  \label{fig:squirm_contour}
\end{figure}

\begin{figure}
  \centering
  \includegraphics[width=.7\columnwidth]{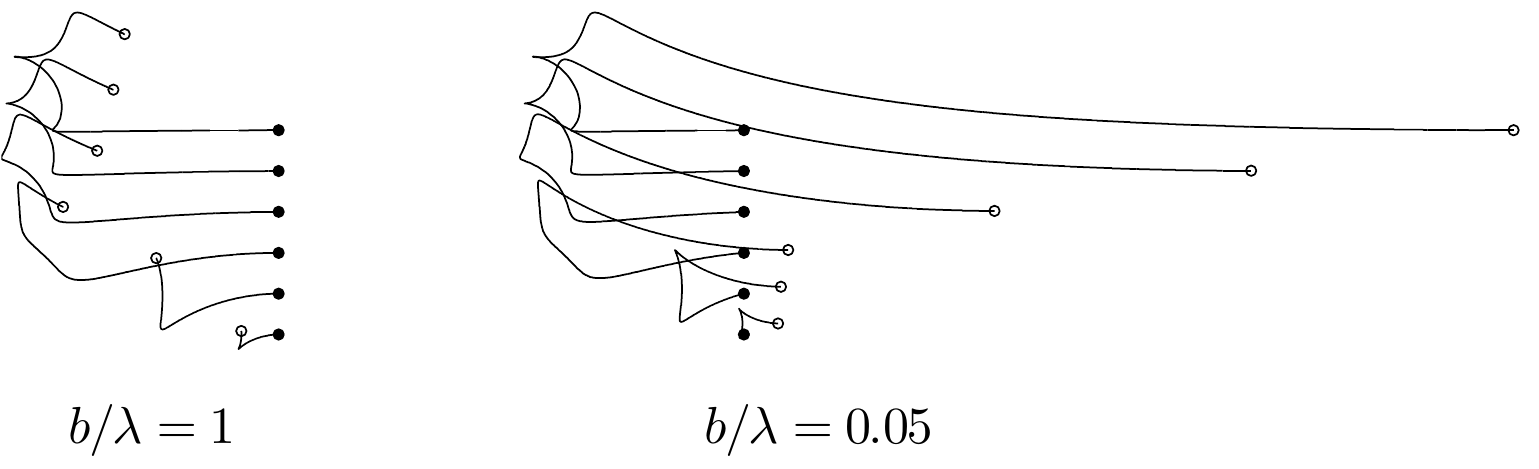}
  \caption{Typical trajectories of a target particle in the fixed lab
    frame for different values of $\ai$ and $\bi$ due to a single
    squirmer moving from left to right, with $\lsc=U=1$, $\pal=100$
    and $\beta=5$.  The trajectories are offset vertically for
    clarity.  The initial positions of the particle are marked by
    solid dots and the final positions by hollow dots. In each panel,
    solid dots from top to bottom correspond to
    $\log(\ai/\lsc)=2,1,0,-1,-2,-3$.}
    \label{fig:squim_onetraj}
\end{figure}

Following~\cite{Ishikawa2006}, we choose a specific instance of an
axially-symmetric squirmer of radius~$\lsc$, swimming along the
positive $z$-axis at a constant speed~$\Uc$. Its free-space, steady,
axisymmetric streamfunction in a comoving reference frame is
\begin{equation}
  \psi(\rho,z)
  = \tfrac12{\rho^2 \Uc}\l(-1 + \frac{\lsc^3}{r^3}
  + \frac{3\beta \lsc^2 z}{2 r^3}\l(\frac{\lsc^2}{r^2} - 1\r)\r)
  \label{eq:squirm_strfcn}
\end{equation}
in cylindrical coordinates where $r = \sqrt{\rho^2+z^2} =
\sqrt{x^2+y^2+z^2}$.  The stresslet coefficient $\beta$ is a free,
dimensionless parameter and the special case $\beta=0$ corresponds to
a sphere in potential flow; we shall take~$\beta=5$ in numerical
examples as in~\cite{Ishikawa2006}.  Figure~\ref{fig:squirm_contour}
shows the contours of the streamfunction~\eqref{eq:squirm_strfcn} for
a squirmer swimming from left to right, and
figure~\ref{fig:squim_onetraj} shows typical trajectories of a passive
target particle displaced by the squirmer with different values of the
impact parameters~$\ai$, $\bi$.  These types of trajectories were
discussed by \citet{Dunkel2010} in the context of swimming organisms.

Compared with the fore-aft symmetry of the potential flow
case~$\beta=0$, a squirmer with~$\beta>0$ has a downstream wake or
`bubble', which we will discuss in section~\ref{sec:wake}.  In
addition, a passive particle displaced by the squirming motion
experiences a much larger excursion when the swimmer is still more
than several body lengths away from the point of initial closest
approach, while in potential flow this contribution is negligible to
the total displacement.  This suggests that the finite-path
displacement function~$\Delta_\pal(\ai,\bi)$ due to a squirmer has a
more complicated dependence on the details of the swimming path
through the values of $\ai$, $\bi$, and $\pal$.  A direct consequence
of this feature is that the diffusivity integral can no longer be
reduced to one dimension by restricting the integration domain of
$\bi$ as in equation~\eqref{eq:Deltahat}.  Instead, the full
two-dimensional integral~\eqref{eq:main3d} has to be computed.

\begin{figure}
  \centering
  \subfigure[]{
   \includegraphics[width=.45\columnwidth]{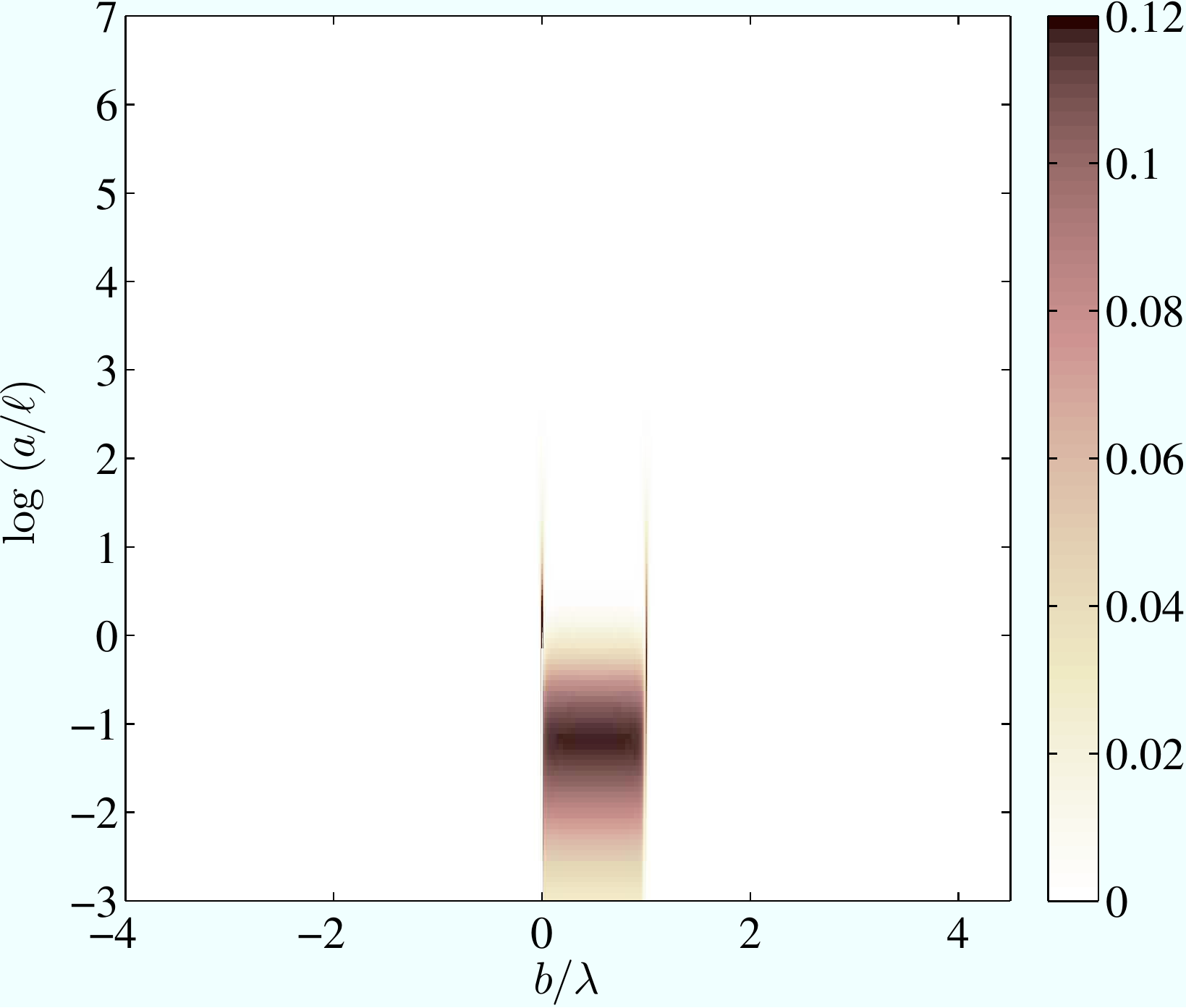}
   \label{fig:sphere_dispint}
   }
  \subfigure[]{
   \includegraphics[width=.43\columnwidth]{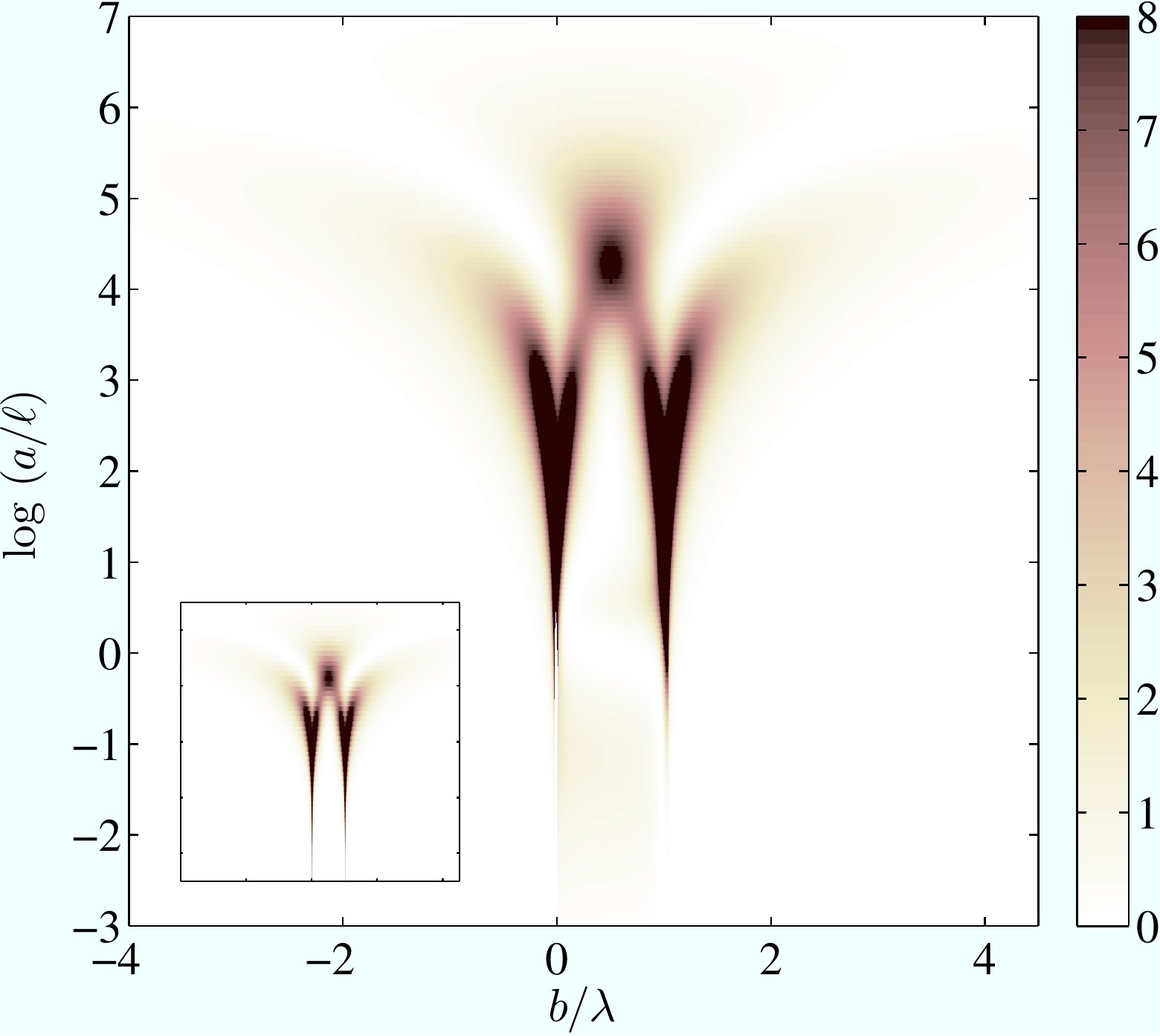}
   \label{fig:squirmer_dispint_ff_inset}
  }
  \caption{Integrand $\lsc^{-4}\ai^2\Delta^2_\pal(\ai,\bi)$
    from~\eqref{eq:main3d_tr} due to (a) potential flow past a sphere;
    (b) a squirmer with $\beta=5$; the near-field potential-flow
    pattern can be dimly seen for~$\log\ai<0$, but is much weaker than
    the far field contribution.  Inset: integrand corresponding to the
    far-field approximation of the squirmer flow.  In all figures,
    $\lsc=\Uc=1$ and $\pal=100$.}
  \label{fig:squirm_vs_sphere_delta2}
\end{figure}

\cite{Thiffeault2010b} showed that the
integrals~\eqref{eq:main2d} and~\eqref{eq:main3d} for potential
swimmers are dominated by ``head-on'' collisions, namely, significant
particle displacement only occurs when the impact parameter $\ai
\lesssim \lsc$ and when the swimmer passes by the particle.  We next
investigate the dominant contribution of the integral for the
squirmers~\eqref{eq:main3d}.  Transforming the integral as
\begin{equation}
  \pal^{-1}\lsc^{-4} \int_{-\infty}^{\infty} \int_0^\infty
  \ai\Delta_\pal^2(\ai,\bi)\dadb =
  \int_{\mathds{R}^2}\lsc^{-4}\ai^2\Delta_\pal(\ai,\bi)
  \dint\log(\ai/\lsc)\dint(\bi/\pal),
\label{eq:main3d_tr}
\end{equation}
we compare in figure~\ref{fig:squirm_vs_sphere_delta2} the spatial
distributions of the dimensionless integrand
$\lsc^{-4}\ai^2\Delta_\pal(\ai,\bi)$ due to a sphere in potential flow
and a squirmer in Stokes flow, for $\lsc=\Uc=1$ and $\pal=100$, in the
$\log (\ai/\lsc)$--$(\bi/\pal)$ plane.  The far-field stresslet-only
approximation to the displacement for a squirmer is also plotted as an
inset in figure~\ref{fig:squirmer_dispint_ff_inset}.  The inset plot
is almost identical to the full plot for any $\pal\gg \lsc$.  Note
that there is a vertical strip near $\bi=0$ in each figure which
corresponds to the situation where the initial particle position is
inside the swimmer or its wake and is therefore excluded in the
calculations.  We shall return to the effect of particles trapped in
the wake in section~\ref{sec:wake}.

For a sphere in potential flow, the dominant contributions
to~\eqref{eq:main3d_tr} come from a rectangle
(figure~\ref{fig:sphere_dispint}) where the values of the integrand
are nearly uniform in $\bi$ for $\bi\in[0,\pal]$.  This justifies the
approximation~\eqref{eq:Deltahat}.  However, for a squirmer the
dominant contributions to the integral~\eqref{eq:main3d_tr} come from
a much larger and irregular region with~$a/\lsc\gtrsim 1$.  The
contribution from this region is negligible in the potential case.

What explains the strong peaks in
figure~\ref{fig:squirmer_dispint_ff_inset}
\begin{figure}
  \centering
   \includegraphics[width=.85\columnwidth]{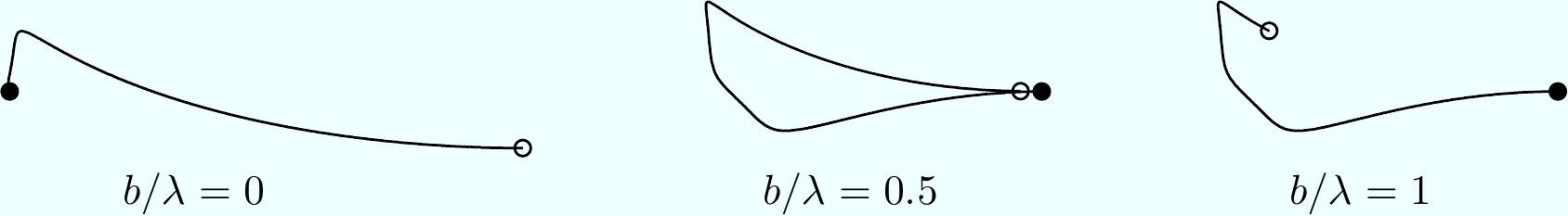}
   \caption{Trajectories of target particle for~$\ai/\lsc=1.1$ and
     different values of~$\bi$.  The filled dot is the initial
     position of the target particle and the hollow dot is its final
     position after the swimmer has moved by a distance~$\pal=100$ to
     the right.  The net displacement is small at~$\bi=\pal/2$ because
     of cancellation as the particle moves towards and then follows
     the swimmer.}
  \label{fig:onetraj_squirmer_a=1.1}
\end{figure}
along the lines~$\bi/\pal=0$, $\bi/\pal=1$?
Figure~\ref{fig:onetraj_squirmer_a=1.1} provides the answer: a target
particle with~$\bi/\pal=1/2$ undergoes a sizable excursion, but has a
small net displacement since it moves towards and then follows the
squirmer, giving back a large part of the displacement it first
underwent.  In contrast, for~$\bi/\pal=0$ and $\bi/\pal=1$ the
squirmer starts or stops (i.e., changes direction) before the particle
reverses direction and gives back its displacement.

\begin{figure}
  \centering
  \subfigure[]{
    \includegraphics[height=0.23\textheight]{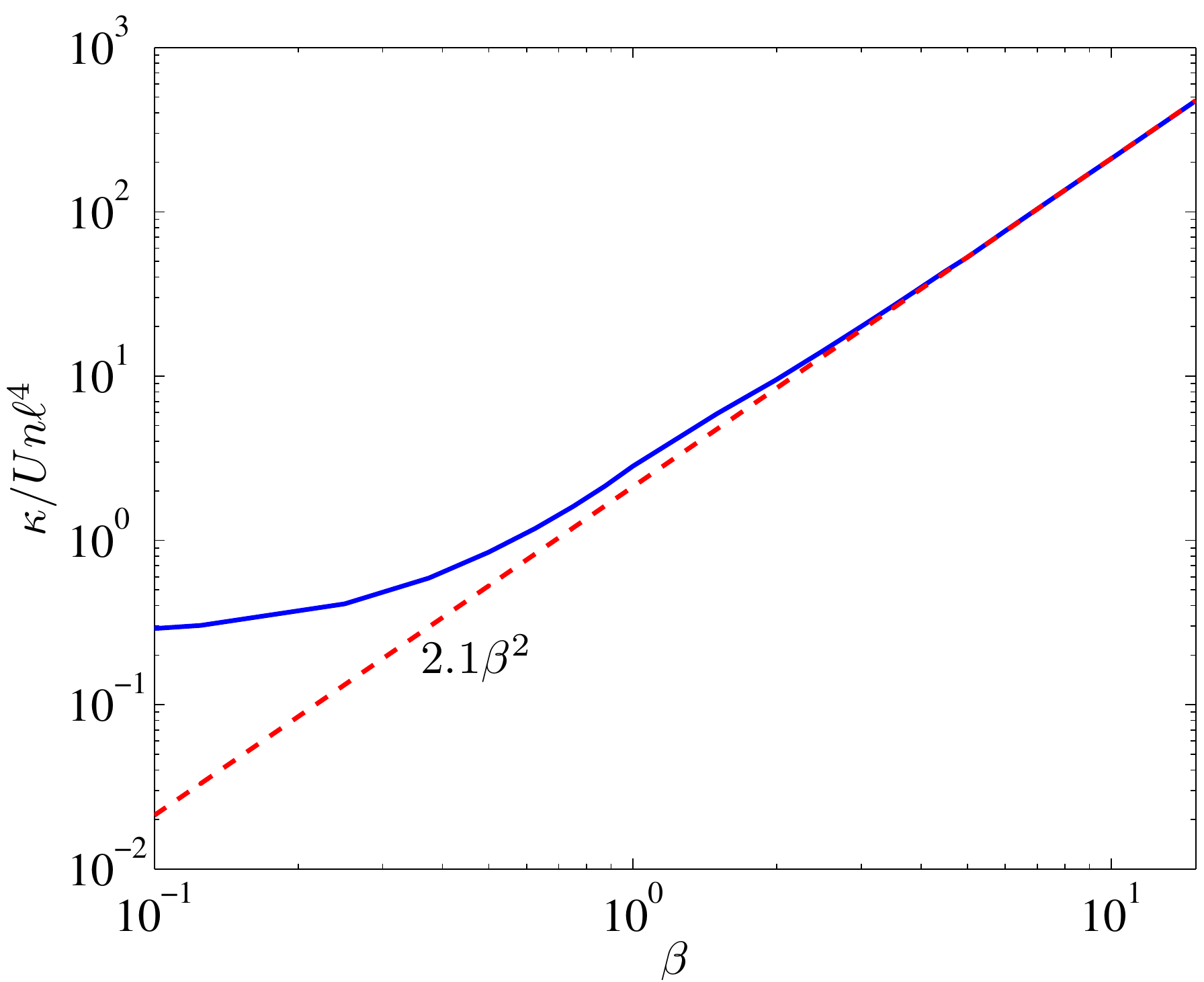}
    \label{fig:squim_kappa_vs_beta}
  }
  \hskip 0.5cm
  \subfigure[]{
    \includegraphics[height=0.23\textheight]{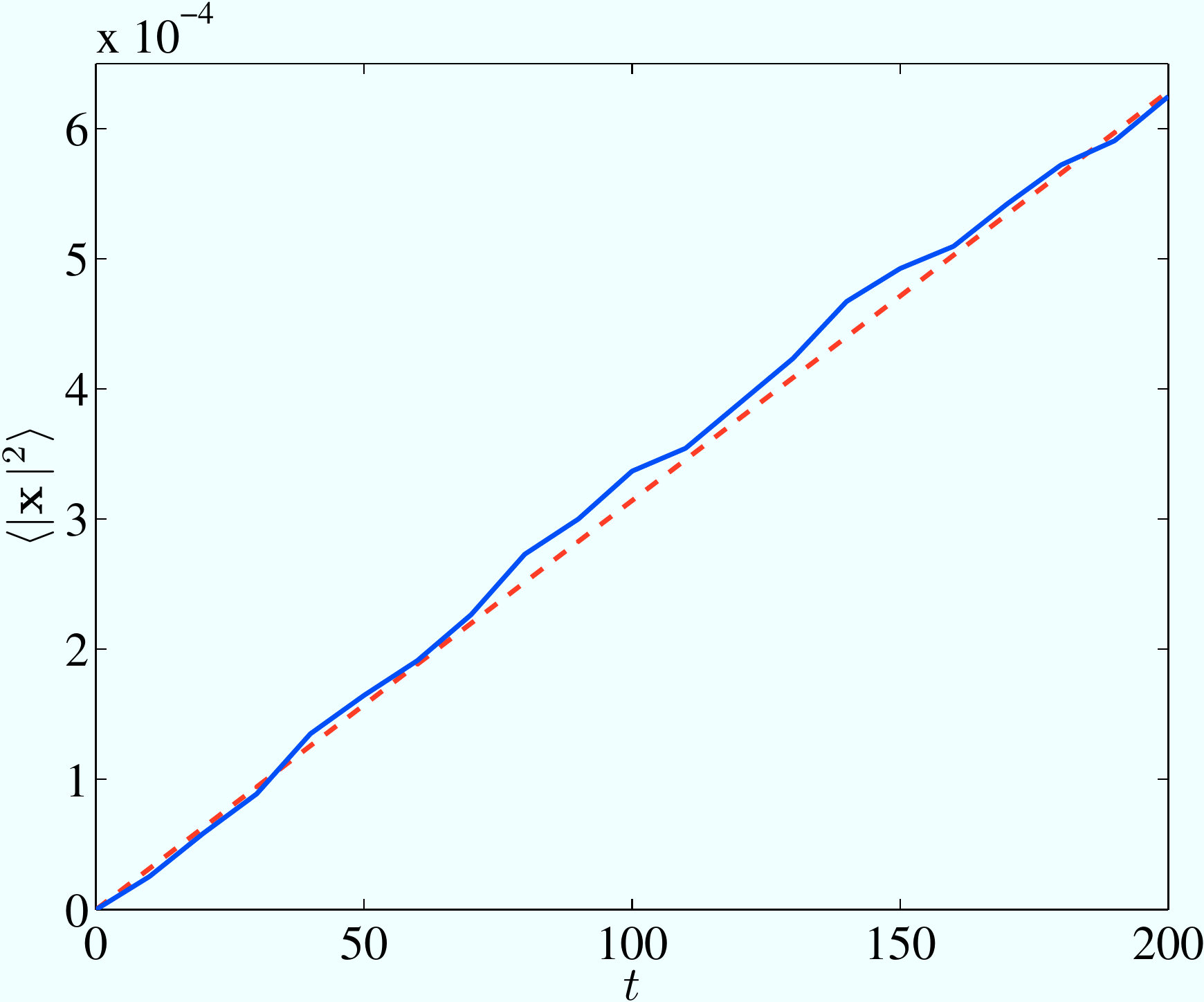}
    \label{fig:trials_single}
  }
  \caption{(a) Dependence of effective diffusivity on $\beta$ for
    $\pal\gg\lsc$: the solid line is the numerical integration
    of~\eqref{eq:main3d} with $\Delta_\pal$ computed with the
    streamfunction~\eqref{eq:squirm_strfcn}.  For large~$\pal$,
    $\kappa$ is independent of~$\pal$.  The dashed line is the
    effective diffusivity due only to the stresslet term.  (b) The
    mean-squared displacement (solid line) of a target particle for
    $2\times 10^6$ realisations of $10$ squirmers, with
    $\lsc=\Uc=1$. The dashed line is the squared-displacement
    predicted by~\eqref{eq:diffsquirm}, using a number
    density~$\nd=10^{-8}\lsc^{-3}$.}
\end{figure}

Now that we know where the contributions to~\eqref{eq:main3d_tr} arise
from, we apply the basic theory to the squirmer model and find
\begin{equation}
 \kappa= 57.0\,\Uc \nd\lsc^4,\quad \pal\gg \lsc
  \label{eq:diffsquirm}
\end{equation}
which is more than $200$ times bigger than the effective diffusivity
due to potential swimmers under the same set of parameters.  One
source of this large enhancement comes from the free
parameter~$\beta$, equal to~$5$ here. Since at far field the dominant
perturbation to the uniform flow in the stream
function~\eqref{eq:squirm_strfcn} is multiplied by~$\beta$, we expect
that for large~$\beta$, $\kappa \approx 2.1\beta^2\Uc \nd\lsc^4$ (the
far-field value, with coefficient fitted in
figure~\ref{fig:squim_kappa_vs_beta}), whereas the diffusivity
converges to the potential value $0.266\,\Uc\nd\lsc^4$ as $\beta\to
0$.  Figure~\ref{fig:squim_kappa_vs_beta} shows how different values
of $\beta$ change the effective diffusivity $\kappa$.
Using the far-field form of our model, we can recover the diffusivity
$1.53\; \mu\mathrm{m}^2/\mathrm{sec}$ measured by \cite{Leptos2009},
for volume fraction~$1.6\;\%$, velocity~$\Uc =
100\;\mu\mathrm{m}/\mathrm{sec}$, and size~$\lsc = 5\;\mu\mathrm{m}$,
using a stresslet amplitude $\beta=0.6$.  The linearity of the
diffusivity in the velocity and volume fraction (for dilute
suspensions) has been pointed out by \cite{Underhill2008} and
\cite{Leptos2009}, but is verified here from first principles with
computed prefactors.

Next we validate the results with direct simulations conducted as
in~\cite{Thiffeault2010b}.  For each trial, we use~$10$
randomly-distributed, non-interacting squirmers with~$\lsc=\Uc=1$
moving in random initial directions in a large periodic box of
size~$1000$.  Each squirmer moves in a straight line for a
length~$\pal=100$ and then turns in a new random direction.  The
turning times are randomized so that the squirmers do \emph{not} all
turn simultaneously.  The displacement statistics are identical when
this asynchrony is relaxed. The target particle initially at the
origin is displaced by the superposition of the flows created by each
squirmer (this is a good approximation at such low densities).
Figure~\ref{fig:trials_single} shows the mean-squared
displacement~$\<\lvert\xv\rvert^2\>$ of a target particle over
$2\times 10^6$ trials (realisations).  The agreement between the solid
curve (numerically-simulated $\<\lvert\xv\rvert^2\>$) and the dashed
line ($6\kappa t$ with $\kappa$ computed with the
integral~\eqref{eq:main3d}) verifies the the theoretical prediction.
We also simulated a large number of particles advected by a single
realisation of~$30$ squirmers: the squared-displacement averaged over
all the particles obeys the theory for large times, as expected.

\section{Wake transport}
\label{sec:wake}

In the previous section we have ignored the wake or closed `bubble'
behind the squirmer, visible in figure~\ref{fig:squirm_contour}, since
it is helpful to treat it separately.  Assume that a particle is
initially trapped in the swimmer's wake: it will then be displaced by
the full swimmer's path length~$\pal$, until the swimmer changes
direction and new particles are introduced in the wake.  We can thus
approximate the displacement~$\Delta_\pal$ to be equal to~$\pal$
if the particle is inside the wake.  The wake contribution to the
effective diffusivity is then obtained from~\eqref{eq:main3d} as
\begin{equation}
  \kappa_\wake
  = \frac{\Uc\nd}{6\pal}\pal^2\int_{\wake} 2\pi \ai\,\dadb
  = \tfrac16\Uc\nd\Vol_\wake\,\pal
  \label{eq:kappawake}
\end{equation}
where~$\Vol_\wake$ is the volume of the wake `bubble'.  Note
that~$\nd\Vol_\wake$ is the total volume fraction of wake bubbles in
the fluid.  The wake effective diffusivity~\eqref{eq:kappawake} is
potentially much larger than that due to displacements outside the
wake bubble, since~\eqref{eq:kappawake} is directly proportional
to~$\pal$.  In practice, many swimmers do not exhibit a wake bubble,
and even for those that do the time-dependence of the velocity field
near the swimmer and the effect of other swimmers can cause particles
to jump in and out of the wake, which may lower~\eqref{eq:kappawake}
considerably.  Molecular diffusivity will also limit the path length
over which material remains trapped in the wake: for $\pal \gtrsim
\Uc\Vol_\wake^{2/3}/\kappa_{\text{molecular}}$, significant amounts of
material will have leaked out.

\section{Nonzero Reynolds number}

\begin{figure}
  \centering
     \subfigure[]{
     \includegraphics[width=0.4005\columnwidth]{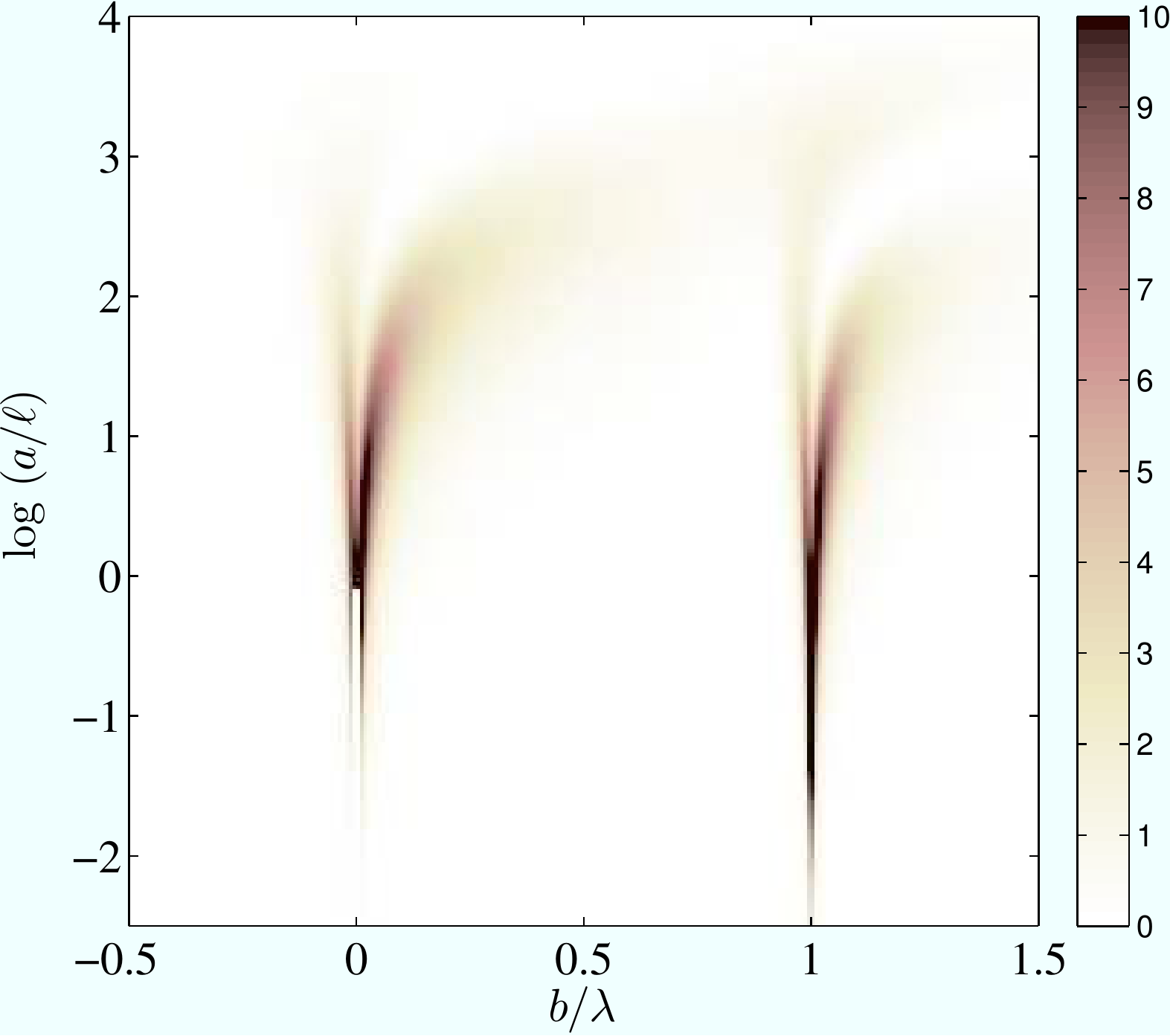}
     \label{fig:oseen_dispint_Re=half}
     }
     \subfigure[]{
     \includegraphics[width=.3600\columnwidth]{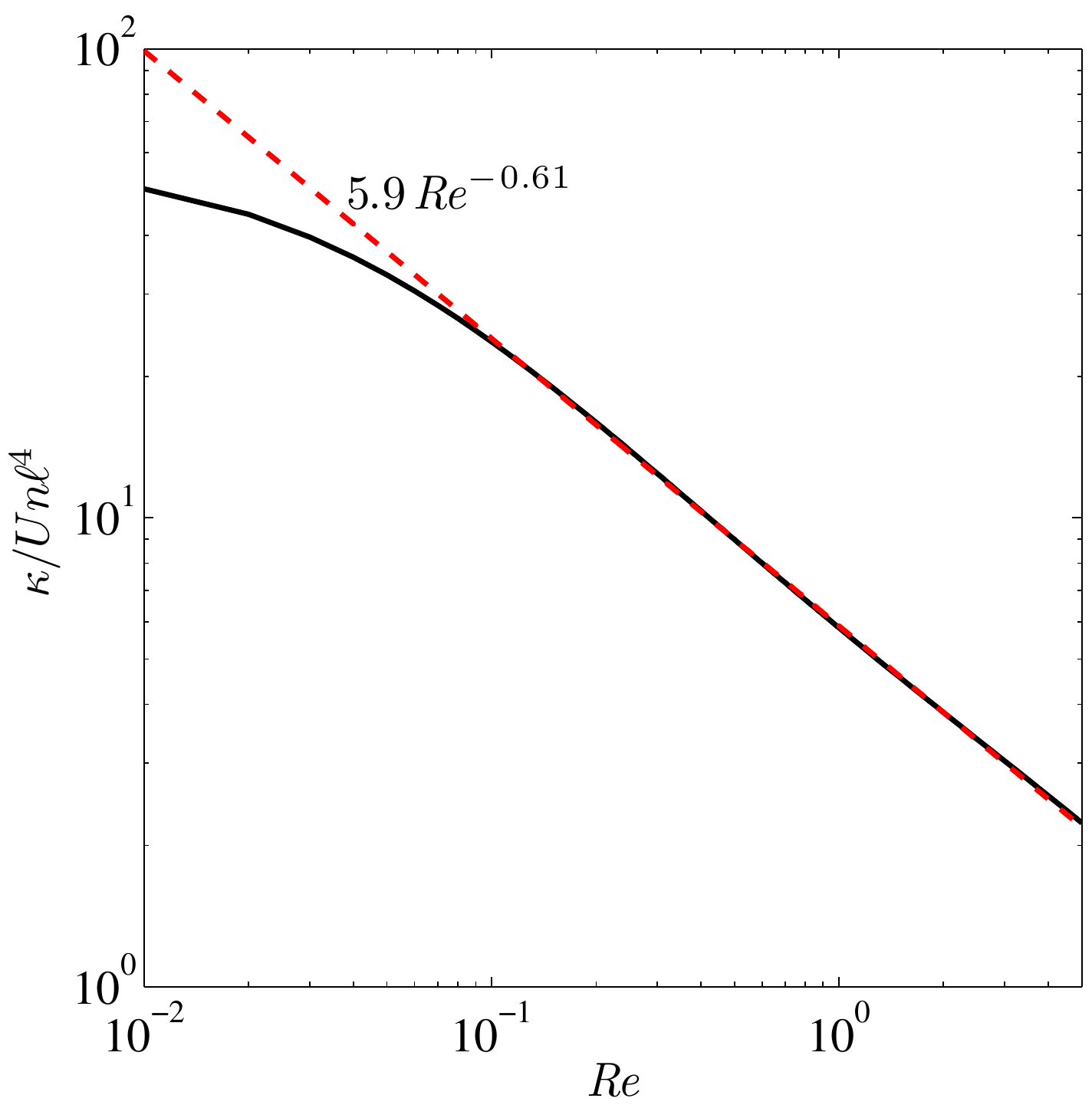}
     \label{fig:oseen_kappa_vs_re}
     }
     \caption{The role of the Reynolds number $\Rey$.  (a) Integrand
       $\lsc^{-4}\ai^2\Delta^2_\pal(\ai,\bi)$
       from~\eqref{eq:main3d_tr} for displacements due to the
       far-field Oseen flow~\eqref{eq:psiOseen}, for~$\Rey=0.5$
       and~$\pal=100$.  (b) The effective diffusivity~$\kappa$
       from~\eqref{eq:main3d} as a function of~$\Rey$. For
       small~$\Rey$ we recover~\eqref{eq:diffsquirm}.}
\end{figure}

In the previous section, the basic theory was applied to model
squirmer-induced mixing in Stokes flow, for which the Reynolds
number~$\Rey=0$ and inertial effects are neglected.  We identified the
dominant contributions as arising from the far field (stresslet).  To
generalise the squirmer far field and to account for finite (but
small) Reynolds number effects on diffusion, we consider the analogous
problem for the Oseen equations, which linearise the inertial
forces~\citep{Oseen1910}.  The singularity analogous to the squirmer
far-field has streamfunction
\begin{equation}
  \psi =
  -\tfrac12{\Uc\rho^2} +
  \frac{\Uc\lsc^2\rho^2}{r^2}
  \l(e^{-\Rey\,(r-z)/2\lsc} +
  \frac{2}{\Rey}\frac{\lsc}{r}\l(e^{-\Rey\,(r-z)/2\lsc}-1\r)\r).
  \label{eq:psiOseen}
\end{equation}
It is easily checked that as $\Rey \to 0$ this recovers the far-field
displacement due to a Stokesian squirmer.  Using this streamfunction
we obtain numerically~$\Delta_\pal(\ai,\bi)$ for different values of
$\Rey$ and summarise the results for the effective diffusivity in
figure~\ref{fig:oseen_kappa_vs_re}.  For small~$\Rey$, we recover the
Stokes value of the effective diffusivity; for larger values
of~$\Rey$, the effective diffusivity drops off as~$\Rey^{-0.61}$ due
to smaller particle displacements.

\section{PDF of displacements}

We close with some comments regarding the typical size of
displacements imparted by the squirmers.  In a recent letter,
\citet{Leptos2009} measured experimentally the probability
distribution function (PDF) of displacements of a target particle due
to repeated `kicks' by swimming micro-organisms.  They observe that,
unlike in Einstein's theory of Brownian motion, the PDF is
non-Gaussian with clear exponential tails.  We now show that squirmers
also exhibit such tails, due to the stagnation points at their
surface.

The large displacements for squirmers are due to particles that lie
very close to the axis of motion, i.e., with~$\ai\ll\lsc$.  For
small~$\ai$, the displacement function can be well-approximated with a
`top-hat' function as in~\eqref{eq:Deltahat}.  The probability
of~$\Delta_\pal(\ai,\bi)$ being larger than a given
value~$\Deltarvar$ is then
\[
P(\Delta_\pal > \Deltarvar)
= \frac{1}{\Vol}\int_{\Delta_\pal(\ai,\bi) > \Deltarvar} \pi\ai\da\db
= \frac{\pal}{\Vol}\int_{\Delta(\ai) > \Deltarvar} \pi\ai\da.
\]
For~$\ai\ll\lsc$, the squirmer displacement~$\Delta(\ai)$ diverges
as~$\C\lsc\log(\lsc/\ai)$, where~$\C$ is a numerical constant.  Hence,
the contribution for large~$\Deltarvar$ comes from small~$\ai$, and we
have
\begin{equation}
  P(\Delta_\pal > \Deltarvar)
  \approx \frac{\pal}{\Vol}\int_{\C\lsc\log\ai^{-1} > \Deltarvar} \pi\ai\da
  = \frac{\pi\pal}{2\Vol}\,\ee^{-2\Deltarvar/\lsc\C},
  \label{eq:exptails}
\end{equation}
which exhibits exponential tails, as in \cite{Leptos2009}.  A simple
argument using large-deviation theory shows that the PDF of a
superposition of such exponential tails retains the same exponential
tails.

\begin{figure}
  \centering
  \subfigure[]{
    \includegraphics[height=0.23\textheight]{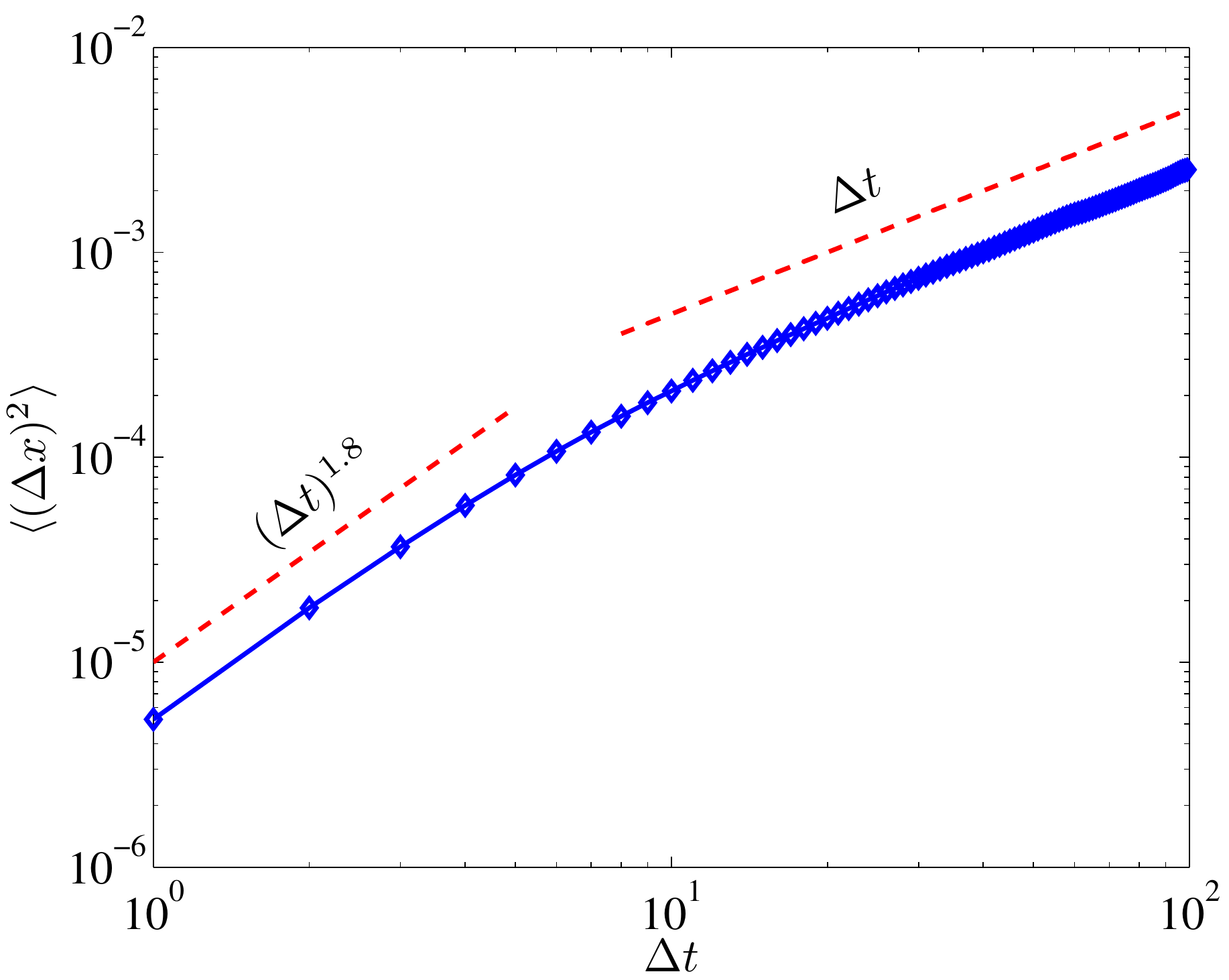}
    \label{fig:leptos_var_scale}
  }
  \hskip 0.5cm
  \subfigure[]{
    \includegraphics[height=0.23\textheight]{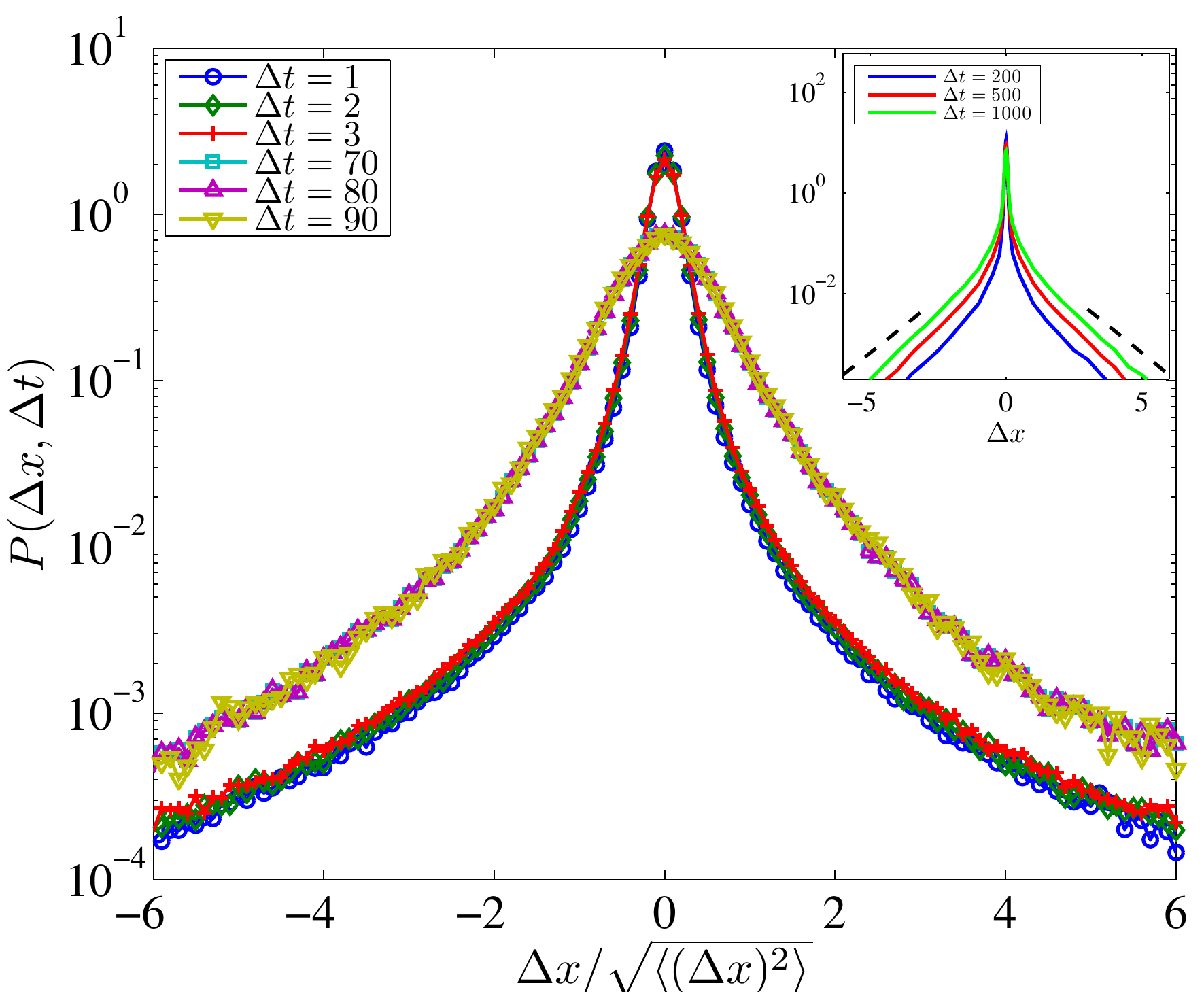}
    \label{fig:leptos_hist}
  }
  \caption{(a) Growth of mean-squared displacement $\langle(\Delta
    x)^2\rangle$ with time, with superdiffusive (short-time) and
    diffusive (long-time) regimes.  (b) Rescaled $x$-displacement
    distributions in the two regimes, showing the collapse after
    scaling.  Inset: unscaled PDFs with exponential
    tails~\eqref{eq:exptails}.}
  \label{fig:leptos}
\end{figure}
To make contact with \citet{Leptos2009}, we statistically analyse
displacements for our biomixing model via particle simulations.  We
examine the displacement histograms for different sampling intervals
$\Delta t$ up to $90$, over which each swimmer travels a distance
$\Uc\Delta t$.  Figure~\ref{fig:leptos_var_scale} shows the
mean-squared displacement $\langle(\Delta x)^2\rangle$ as a function
of time: for short times, this grows as~$(\Delta t)^{1.8}$, in good
accordance with the superdiffusive regime observed by \citet{Wu2000}:
they found an exponent between $1.5$ and $2$.  For longer times, we
recover a linear diffusive regime, as expected, with slope given by
our theoretical prediction~$2\kappa$.  Figure \ref{fig:leptos_hist}
shows that the incremental displacement probability distribution along
the $x$-axis, $P(\Delta x,\Delta t)$, rescales to a single
distribution function in terms of the similarity variable $\Delta
x/\sqrt{\langle(\Delta x)^2\rangle}$ in each of the two scaling
regimes.  However, note that the exponential tails~\eqref{eq:exptails}
are independent of $\Delta t$, and are only visible in simulations
after integrating for a long time (see inset in
figure~\ref{fig:leptos_hist}).  Thus, it is not clear if these are
related the the exponential tails observed by \citet{Leptos2009}.

\begin{acknowledgments}
  The authors are grateful to Keith Moffatt for suggesting the wake
  transport calculation, to Eric Kunze for pointing out that molecular
  diffusion would limit the wake transport, and to the Institute for
  Mathematics and its Applications (supported by NSF) for its
  hospitality.  SC was supported by NSF under grant DMS-0507615, J-LT
  under grant DMS-0806821.
\end{acknowledgments}


\end{document}